\documentclass[12pt]{iopart}
\def\({\left(}
\def\){\right)}
\def\[{\left[}
\def\]{\right]}

\begin{document}
\paper[Renormalization Group and Infinite Algebraic
Structure\ldots]
 {Renormalization Group and Infinite Algebraic
Structure in $D$-Dimensional Conformal  Field Theory. }
\author
{Yu M Pis'mak {\dag}}
\address
 {\dag \ Department of Theoretical Physics,  State University Saint-Petersburg, Russia}
\begin{abstract}
We consider scalar field theory in the
    $D$-dimensional space with nontrivial metric
    and local action functional of most general
    form. It is possible to construct for this
    model a generalization of renormalization
    procedure and  RG-equations. In the fixed
    point the  diffeomorphism and Weyl
    transformations generate an infinite algebraic
    structure of $D$-Dimensional conformal field
    theory models. The Wilson expansion  and crossing symmetry
    enable to obtain sum rules for dimensions of composite
    operators and Wilson coefficients.
    \end{abstract}
\submitto{\JPA} 
\section{Introduction}
Essential achievements in investigations of quantum field
theoretical models were obtained on the basis of analysis of their
symmetry properties and algebraic structures. The higher is the
symmetry, the more restrictions put it on the possible form of
correlation functions. It is known that the conformal invariance
defines two point correlation functions up to a constant amplitude
and three point ones as a finite linear combination of known
functions (Polyakov triangles) with constant coefficients
\cite{P},\cite{TMP},\cite{FP}. In studies of 2-dimensional
conformal field theory it was found that the Virasoro algebra
described its most fundamental features \cite{BPZ},\cite{FMS}. An
approach to the extension of these methods on conformal field
theory in $D$ dimensions was suggested in \cite{DP1} for any $D$.
In this papers, it was proposed to use the algebra of the general
coordinate transformation as an analog of the Virasoro algebra for
the $D$-dimensional case. In \cite{DP1} the Green functions for
operators $\phi^2$, $\phi^4$ were studied in the theory $\phi^4$.
For a generalized diffeomorphism combined from diffeomorphism and
Weyl transformations the Ward identities for these Green functions
were obtained which are similar to ones used in 2-dimension
conformal theory. In this paper we generalize the results of
\cite{DP1} for all the composite operators of the scalar
$D$-dimensional conformal fied theory and construct the infinite
algebraic structure analogous to one presented by Virasoro algebra
in two dimensions. For this purpose we analyze the diffeomorphism
and Weyl transformation in $D$-dimensional curved space and the
most essential features of the renormalization procedure in
quantum field theory \cite{P1}. We use the Wilson operator product
expansion and crossing symmetry  for construction of sum rules for
critical exponents and Wilson coefficients in conformal field
theory \cite{Pp}.

\section{Diffeomorphism transformations}
For the curved $D$-dimensional space with metric
$\gamma_{\mu\nu}$, the general coordinate (diffeomorphism)
transformations are defined in the following way.  The
infinitesimal reparametrization of the coordinates $x$ is written
as : $ \delta_{\alpha}^{DT}x^{\mu} = \alpha^{\mu}(x) $, where
$\alpha_{\mu}(x)$ are the parameters of transformation. The
commutation relation for diffeomorphism transformations (DT) is of
the form:
\begin{equation}
[\delta_{\alpha}^{DT},\delta_{\beta}^{DT}] =
\delta_{[\alpha,\beta]}^{DT}, \label{crdt}
\end{equation}
where $ [\alpha,\beta] = (\alpha\nabla)\beta - (\beta\nabla)\alpha
$ is the commutator of vector fields ($\nabla_\lambda$ denotes the
covariant derivative, $\nabla_{\lambda}\gamma_{\mu\nu} = 0$). For
tensor fields,
\begin{equation}
\delta_{\alpha}^{DT}F(x) =L_{\alpha}F(x) \label{Lid}
\end{equation}

Here $ L_{\alpha} $  denotes the Lie derivative defined by the
vector field $\alpha^{\mu}(x)$:
$$
 L_{\alpha}F_{\nu_1,...,\nu_n}^{\mu_1,...,\mu_m}(x) =
(\alpha\nabla) F_{\nu_1,...\nu_n}^{\mu_{1},...,\mu_{m}}+
\sum_{i=1}^n \nabla_{\nu_i} \alpha^{\lambda_i}
F(x)_{\nu_1,..,\lambda_i,...,\nu_n}^{\mu_1,...,\mu_m}-
\sum_{i=1}^m \nabla_{\lambda_i} \alpha^{\mu_i}
F(x)_{\nu_1,...,\nu_n}^{\mu_1,...,\lambda_i,...,\mu_m}.
$$
Particularly, for the scalar field $\phi$ and the metric
$\gamma_{\mu\nu}(x)$
\begin{equation}
\delta_{\alpha}^{DT}\phi(x) = (\alpha \nabla)\phi(x), \ \
\delta_{\alpha}^{DT}\gamma_{\mu\nu} = \nabla_{\mu}\alpha_{\nu}
 +
\nabla_{\nu}\alpha_{\mu} \label{DTgamma}
\end{equation}

    Let us introduce the notation:
$$
\omega^{\mu\nu}_{\alpha}(\gamma) \equiv \nabla^{\mu}\alpha^{\nu} +
 \nabla^{\nu}\alpha^{\mu} - \frac{2}{D}
 (\nabla\alpha)\gamma^{\mu\nu} .
$$
The vector $\alpha (x)$ for which
 $\omega_{\alpha}^{\mu\nu} = 0$ and the corresponding transformation
$\delta_{\alpha}^{conf} \equiv \delta_{\alpha}^{DT}$ will be
called  conformal ones. It is well known that in the flat space
$\delta_{\alpha}^{conf}x$ is the conformal transformation (CT) of
the coordinates $x$. The commutator $ [\alpha,\beta]$ of conformal
vectors $\alpha$, $\beta$ is conformal. Therefore it follows from
(\ref{crdt}) that the CTs form a subgroup of the DT group. This
subgroup will be called conformal.

Let $\Phi(x)_{\mu_1,...,\mu_n}$ denotes the covariant tensor,
obtained by multiplications of the covariant derivatives of the
field $\phi$ and the curvature tensors with possible contraction
of part of the indices. The set $\{{\bf\Phi}\}$ of all such
tensors  can be used as the basis for constructing diffeomorphism
invariant local functionals of the field $\phi$. Thus the most
general form of the diffeomorphism invariant local action of the
field $\phi$ in the curved D-dimensional space can be written as
follows [3,4]: $$
 S(A,\gamma, \phi) = \int
dx \sqrt{\gamma}L(\phi(x),A(x)), $$
 where $\gamma \equiv \det
\,\gamma_{\mu\nu}$, $L(\phi(x),A(x))$ is the Lagrangian:
 $$
L(\phi(x),A(x),\gamma (x)) = \sum_{\Phi_i\in \{{\bf\Phi}\} }
A^i(x)\Phi_i(x) $$ and $A^i(x)$ denotes the contravariant tensor
source corresponding to the covariant tensor field $\Phi_i(x)$. In
the Lagrangian the indices of sources and the fields are
contracted.

   The generating functional for the connected Green functions of
the Euclidean quantum field theory with action $S$ has the form:
$$
   W(A,\gamma) = \ln \int \exp\{-S(\phi, A)\} D\phi
$$
 The metric $\gamma_{\mu\nu}(x)$ can be considered as the
source for the energy-momentum tensor.

Taking into account the diffeomorphism invariance of the action
$S(\phi, A)$  and using the Schwinger equations for $W$ it is easy
to show that the functional $W(A,\gamma)$  is invariant in respect
to the DTs:
\begin{equation}
 \delta_{\alpha}^{DT}W(A,\gamma) =
D_{\alpha}^{DT} W(A,\gamma) = 0. \label{DTinv}
 \end{equation}
We have used the notation:
 $$
D_{\alpha}^{DT}(A,\gamma) \equiv
\delta_{\alpha}^{DT}\gamma_{\mu\nu}
\frac{\delta}{\delta\gamma_{\mu\nu}} + \sum_i
\delta_{\alpha}^{DT}A^i \frac{\delta}{\delta A^i}
 $$
where $\delta_{\alpha}^{DT}\gamma_{\mu\nu}$,
$\delta_{\alpha}^{DT}A^i$  are defined by (\ref{Lid}),
(\ref{DTgamma}). Obviously, the operators
$D_{\alpha}^{DT}(A,\gamma)$ form a representation of the
diffeomorphism algebra:
 $$
[D_{\alpha}^{DT}(A,\gamma), D_{\beta}^{DT}(A,\gamma)] =
D_{[\alpha,\beta]}^{DT}(A,\gamma).
$$

\section{Weyl transformations}
We consider now the group of the Weyl transformations (WT). For
the metric $\gamma_{\mu\nu}$, the infinitesimal WT is defined as
the local rescaling
\begin{equation}
\delta_{\sigma}^W\gamma_{\mu\nu} (x) =
-2\sigma(x)\gamma_{\mu\nu}(x) \label{WTgamma}
\end{equation}
specified by the scalar function $\sigma(x)$.

These transformations form the commutative algebra:
\begin{equation}
[\delta^W_\sigma,\delta^W_{\rho}] = 0. \label{crw}
\end{equation}
For the field $\phi$ we define the WT in the following way:
\begin{eqnarray}
\delta_{\sigma}^{W}\phi(x) = \sigma(x) d_{\phi}\phi(x),
\label{phiw}
\end{eqnarray}
where $ d_{\phi}=
 (D-2)/2
$ is  the canonical dimension of the field $\phi$.
 The definitions (\ref{WTgamma}), (\ref{phiw}) make it possible to define the WT
for the set of fields $\Phi$:
\begin{eqnarray}
 \delta_{\sigma}^{W}\Phi(x) \equiv \(
\delta_{\sigma}^{W}\gamma_{\mu\nu}
\frac{\delta}{\delta\gamma_{\mu\nu}} + \delta_{\sigma}^{W}\phi
\frac{\delta}{\delta\phi}\) \Phi(x). \nonumber
\end{eqnarray}

  This transformation can be written in the form:
\begin{eqnarray}
\delta_{\sigma}^{W}\Phi_{i}(x)= \sum_{j}
M^{j}_i(\sigma)\Phi_{j}(x) \nonumber
\end{eqnarray}
with the matrix $M^{j}_i(\sigma)=M^{j}_i(\sigma,\gamma)$
satisfying the relation:
\begin{eqnarray}
\delta_{\sigma}^{W}\gamma_{\mu\nu}
\frac{\delta}{\delta\gamma_{\mu\nu}}M(\rho) -
\delta_{\rho}^{W}\gamma_{\mu\nu}
\frac{\delta}{\delta\gamma_{\mu\nu}}M(\sigma) +[M(\sigma),M(\rho)]
= 0.  \label{wrel}
\end{eqnarray}

Let us define the operator
\begin{eqnarray}
 D^{W}_{\sigma}(A,\gamma,\phi) \equiv
\delta_{\sigma}^{W}\gamma_{\mu\nu}
\frac{\delta}{\delta\gamma_{\mu\nu}} + \delta_{\sigma}^{W}\phi
\frac{\delta}{\delta\phi} + \sum_{i} \delta_{\sigma}^{W} A^{i}
\frac{\delta}{\delta A^{i}}, \nonumber
\end{eqnarray}
were
\begin{equation}
\delta_{\sigma}^{W} A^{i} \equiv \sum_{j} \left(\sigma
D\delta^{i}_{j} - M^{i}_{j}(\sigma)\right)A^{j}. \label{wta}
\end{equation}
It can be considered as a general form of infinitesimal WT
suitable for all fields and sources because from (\ref{wrel}),
(\ref{wta}) it follows that
$$
 [D^{W}_{\sigma}(A,\gamma,\phi),
 D^{W}_{\rho}(A,\gamma,\phi)] =0.
$$
For the WT defined in this way  one can easily prove that the
action $S$ is invariant:
\begin{equation}
D^{W}_{\sigma}(A,\gamma,\phi)S(A,\gamma,\phi) =0. \label{WTinv}
\end{equation}

 As for the case of the DTs, it follows from
(\ref{WTinv}) that the functional $W$  is invariant in respect to
the WTs:
\begin{equation}
 D^{W}_{\sigma}(A,\gamma)W =0
\label{Winv1}
\end{equation} where
\begin{eqnarray}
D^{W}_{\sigma}(A,\gamma) \equiv D^{W}_{\sigma}(A,\gamma, 0) =
\delta_{\sigma}^{W}\gamma_{\mu\nu}
\frac{\delta}{\delta\gamma_{\mu\nu}} + \sum_{i}
\delta_{\sigma}^{W} A^{i} \frac{\delta}{\delta A^{i}}.
\label{WTred}
\end{eqnarray}
 The operators $D^{W}_{\sigma}(A,\gamma)$ form a
representation of the WT algebra:
$$
 [D^{W}_{\sigma}(A,\gamma),
 D^{W}_{\rho}(A,\gamma)] =0.
$$

For a constant $\sigma$,
\begin{eqnarray}
 \delta_{\sigma}^{W} \Phi_{i}=
d_{i}\sigma \Phi_{i}, \ \ M^i_j(\sigma)=\delta^i_j\sigma d_j, \ \
\delta_{\sigma}^{W} A^{i}= {\bar d_{i}}\sigma A^{i}, \nonumber
\end{eqnarray}
where the constant parameters $d_{i}=d_{i}(D) $,
${\bar d_{i}}\equiv D-d_{i} ={\bar d_{i}}(D)$
 are the dimensions
of the field $\Phi_{i}$ and source $A^{i}$. For the field
$\Phi_{0} \equiv \nabla_{\mu}\phi\nabla^{\mu}\phi$, $ d_{0}=D $
and for corresponding source $A^{0}$, $\bar{d}_{0} = 0$. If the
source $A^{i}$ is dimensionless, i.e.
 $\bar{d}_{i}(D)=0$,
for some definite value $D={\cal D}_i$ of the space dimension,
this dimension ${\cal D}_i$ is called logarithmic for  $A_{i}$.
For given ${\cal D}$ we denote the dimensions of fields and
sources as $ \bar{d}_j^{log}=\bar{d}_j|_{D={\cal D}},\
\bar{d}_j^{log}=\bar{d}_j|_{D={\cal D}}$.

\section{Renormalization}
To perform the renormalization procedure we choose the source $A$
that defines the logarithmic dimension of space $ {\cal D}$ which
is considered as a fixed parameter specifying the renormalized
theory. The generating functional of renormalized Green functions
$W_{r}$ is defined as follows:
\begin{eqnarray}
W_{r}(J,\gamma)  \equiv W(A(J,\gamma),\gamma). \nonumber
\end{eqnarray}
The functions $ A(J,\gamma) $ in the right hand side are of the
form:
\begin{equation}
A^{i} = \mu^{\Delta_{i}}F^{i}(J,\gamma, D). \label{rt}
\end{equation}
Here $\mu$ is an auxiliary scaling parameter ,
\begin{eqnarray}
\Delta_{i}=\Delta_{i}(D) = {\bar d_{i}}(D)
 - {\bar d_{i}}^{log},  \ \  \frac{\partial F^i(J,\gamma,D)}{\partial J^i}\Bigg|_{J=0}=1.
\nonumber
\end{eqnarray}
The function $F(J,\gamma,D)$  obeys  the homogeneity condition
\begin{eqnarray}
D^{log}(J,\gamma)F^i(J,\gamma,D)=\bar{d}_i^{\log}F^i(J,\gamma,D),
\ \ \label{hc}
\end{eqnarray}
where
\begin{eqnarray}
D^{log}(J,\gamma)\equiv
  \sum_{i}\bar{d}^{log}_i
J^{i} \frac{\delta}{\delta J^{i}}
+2\gamma_{\mu\nu}\frac{\delta}{\delta\gamma_{\mu\nu}}.
\label{Dlog}
\end{eqnarray}
It is supposed also that the  functions $ J(A,\gamma) $ defined by
(\ref{rt}) are the tensors with respect to the DTs:
\begin{equation}
\delta_{\alpha}^{DT}J^i =L_{\alpha}J^i. \label{rdt}
\end{equation}

     The operators $D_{\alpha}^{DT}(A,\gamma)$,
$D^{W}_{\sigma}(A,\gamma)$ can be represented in terms of the
variables of $ W_{r} $. It follows from (\ref{WTred}), (\ref{rdt})
that
\begin{equation}
D_{\alpha}^{DT}(A,\gamma) \equiv {\cal D}_{\alpha}^{DT}(J,\gamma)
= D_{\alpha}^{DT}(J,\gamma) , \label{DTr}
\end{equation}
\begin{equation}
D^{W}_{\sigma}(A,\gamma) \equiv {\cal D}^{W}_{\sigma}(J,\gamma) =
\delta_{\sigma}^{W}\gamma_{\mu\nu}
\frac{\delta}{\delta\gamma_{\mu\nu}} + \sum_{i}
\delta_{\sigma}^{W} J^{i} \frac{\delta}{\delta J^{i}}. \label{Wr}
\end{equation}
   The WT for the sources $J$  can be obtained from
(\ref{wta}), (\ref{WTred}),  (\ref{rt}):
 $$
\delta_{\sigma}^{W} J^{i} (J,\gamma) =
\sum_jT^{i}_{j}\left\{\sum_k\left[\sigma D\delta^{j}_{k} -
M^{j}_{k}(\sigma)\right]F^{k}- 2\sigma\gamma_{\mu\nu}
\frac{\delta}{\delta\gamma_{\mu\nu}} F^j\right\}.
 $$
Here $ T^{i}_{j} $ is the element of the matrix $T$ defined as
follows:
 $$
\sum_{j} T^{i}_{j}\frac{\partial F^{j}}{\partial
J^{k}}=\delta^i_k.
 $$
Since the commutation relations are independent of the choose of
variables,
\begin{equation}
[{\cal D}_{\alpha}^{DT}(J,\gamma), {\cal
D}_{\beta}^{DT}(J,\gamma)] = {\cal
D}_{[\alpha,\beta]}^{DT}(J,\gamma), \label{crdt1}
\end{equation}
\begin{equation}
 [{\cal D}^{W}_{\sigma}(J,\gamma),
 {\cal D}^{W}_{\rho}(J,\gamma)] =0.
\label{crw1}
\end{equation}
 In virtue of (\ref{rt}), (\ref{hc}), (\ref{Dlog}), we obtain
 $D^{log}(J,\gamma)=D^{log}(A,\gamma)$. Hence, for constant $\sigma$
it holds:
\begin{eqnarray}
D_{\sigma}^{W}(A,\gamma)= \sigma D^{log}(A,\gamma)+\sigma
\sum_{i}\Delta_iA_i\frac{\delta}{\delta A_i}= \nonumber
\\
=\sigma D^{log}(J,\gamma)+\sigma \sum_{i}\mu\frac{\partial
A_i}{\partial \mu}\Bigg|_{J=const}\frac{\delta}{\delta A_i}=
\nonumber
\\
=\sigma D^{log}(J,\gamma)+\sigma\mu\frac{\partial }{\partial
\mu}\Bigg|_{J=const}-\sigma\mu\frac{\partial }{\partial
\mu}\Bigg|_{A=const},\nonumber
\end{eqnarray}
and
\begin{eqnarray}
\delta^W_\sigma J^i= D_{\sigma}^{W}(A,\gamma)J^i=
\sigma\left(\bar{d}_i^{log} J^i- \mu\frac{\partial J^i}{\partial
\mu}\Bigg|_{A=const}\right). \label{Wrg}
\end{eqnarray}

It follows from (\ref{DTinv}), (\ref{Winv1}) that $W_r(J,\gamma)$
is invariant in respect to the diffeomorphism and Weyl
transformations:
\begin{equation}
 {\cal D}^{DT}_{\alpha}(J,\gamma)W_r(J,\gamma) =0,
\label{DTinv1}
\end{equation}
\begin{equation}
 {\cal D}^{W}_{\sigma}(J,\gamma)W_r(J,\gamma) =0.
\label{Winv2}
\end{equation}

   The functional $W_{r}(\lambda_{r}, J_{r}, J_{\gamma})$ for
usual models of the quantum field theory in Euclidean
$D$-dimensional space can be constructed from $W_r(J,\gamma)$ in
the following way:
$$
 W_{r}(\lambda_{r}, J_{r}, J_{\gamma}) =  W_{r}(\lambda_{r}
+ J_{r}, \gamma^{E} + J_{\gamma}).
 $$
Here $\lambda_{r}$,$J_{r}$ denote the set of renormalized
parameters of the model and the set of the sources of renormalized
composite operators.The source of the energy-momentum tensor and
the metric of $D$-dimensional Euqlidean space are denoted by
$J_{\gamma}$, $ \gamma^{E} $. If  $\lambda_i\neq 0$ only in the
case $\bar{d}^{log}_i\geq 0$, then the model is renormalizable. By
appropriate choosing of the functions $F^i$ in (\ref{rt}) for the
renormalizable model, the functional $W_r(\lambda_r, J_r,
J_{\gamma})$ and the operators (\ref{DTr}),(\ref{Wr}) are finite
for $J_i\rightarrow J_i+\gamma_i$, $\gamma\rightarrow
\gamma^{E}+J_\gamma$, $D={\cal D} $, finite parameters $\lambda_r$
and sources $J_r, J_\gamma $ \cite{BC}, \cite{BD}, \cite{C}. For
$W_{r}(\lambda_{r}, J_{r}, J_{\gamma})$ the equation
(\ref{Winv2}) by constant $\sigma$  appears to be the usual
renormalization group equation, if one takes into account
(\ref{Wrg}) and (\ref{DTinv1}) with $\alpha(x)=x\sigma$.

\section{Critical point }
Combining the DT and the WT one can obtain the transformation of
the form: $$
 \delta_{\alpha} =  \delta_{\alpha}^{DT} +
 \left.\delta_{\sigma}^W\right|_{\sigma =
\frac{\nabla\alpha}{D}}.
 $$
As a consequence of the commutation relations (\ref{crdt}),
(\ref{crw}), it follows that $$
 [\delta_{\alpha},\delta_{\beta}] =
\delta_{[\alpha,\beta]}.
 $$

This means, that the transformations $\delta_{\alpha}$ form the
representation of the diffeomorphism algebra. In virtue of
(\ref{DTgamma}), (\ref{WTgamma}),
 $$
\delta_{\alpha}\gamma^{\mu\nu} = - \omega^{\mu\nu}_{\alpha}.
 $$
Hence, $\delta_{\alpha}\gamma^{\mu\nu}=0 $ for conformal $\alpha$.
Let us introduce the notations: $$
 {\cal D}_{\alpha}(J,\gamma)
\equiv {\cal D}_{\alpha}^{DT}
(J,\gamma) + {\cal
D}_{\sigma}^{W}(J,\gamma) \mid_{\sigma = \frac{\nabla\alpha}{D}},
$$
$$
 {\cal D}_{\alpha}^{r}(\lambda_r, J_r, J_\gamma) \equiv {\cal
D}_{\alpha}(\lambda_{r} + J_r, \gamma^E + J_\gamma ).
$$

It follows from (\ref{crdt1}), (\ref{crw1}) that
$$
 [{\cal D}_{\alpha}^r (\lambda_r,
J_r, J_\gamma), {\cal D}_{\beta}^r (\lambda_r, J_r, J_\gamma)] =
 {\cal D}_{[\alpha,\beta]}^r(\lambda_r J_r, J_\gamma),
$$
 i.e. the operators  $D_{\alpha}^r$ form the representation of the
diffeomorphism algebra. In virtue of the diffeomorphism and Weyl
invariance of $W_r$,
\begin{equation}
{\cal D}_{\alpha}^{r}(\lambda_{r}, J_{r}, J_\gamma)
W_{r}(\lambda_{r}, J_{r}, J_\gamma)=0. \label{Wed}
\end{equation}
If
$$
 {\cal D}_{\alpha}^{r}(\lambda^{*}, 0, 0) = -
\omega^{\mu\nu}_{\alpha}(\gamma^E) \frac{\delta}{\delta
\gamma^{\mu\nu }}
$$
 for parameters $\lambda_r = \lambda^*$ of
renormalized Euclidean theory, we call this set of parameters the
critical point. It can be proved that this equality is equivalent
to equalities defining the fixed point \cite{C} in the
renormalization group theory.

    Let us denote
${\cal D}^{conf}_{\alpha}(J_r,J_\gamma) \equiv {\cal
D}_{\alpha}^{r}(\lambda^*, J_r, J_{\gamma})$ for conformal
$\alpha$.
 For the flat space ,
it follows from (\ref{Wed}) that
\begin{equation}
{\cal D}^{conf}_{\alpha}(J_r, J_\gamma)W_{r}(\lambda^*, J_r,
J_{\gamma}) = 0. \label{Confinv}
\end{equation}
In virtue of ${\cal D}^{conf}_{\alpha}(0,0) = 0$ the equality
(\ref{Confinv}) means the common conformal invariance of Euclidean
quantum field theory at the critical point.

\section{Wilson expansion}
The obtained infinite number of Ward identities presenting the
investigated algebraic structure of conformal field theory are
lineare differential equations for $W_r$ in variation derivatives
of first order. The well known Wilson asymptotic expansion is
written as a differential relation including variation derivatives
of second order:
\begin{eqnarray}
  \frac{\delta}{\delta J^i (x)} \frac{\delta}{\delta J^k (y)}W_r =
   \sum_l\int dz K_{ijl}(x,y,z) \frac{\delta}{\delta J^l
   (z)}W_r.
\label{WE}
\end{eqnarray}
It can be considered as a completing  condition for considered
algebraic structure. In conformal field theory the
Wilson-expansion series are convergent \cite{LM}, the functions
$K_{ijl}(x,y,z)$ being 3-point correlation function are defined
exactly up to finite number of constant (Wilson coefficients) by
dimensions of field operators. In this paper we study restrictions
following from (\ref{WE}) for dimensions of fields and Wilson
coefficients. For this purpose we introduce some definitions and
notations.

Let $x^{(n)}$ be a symmetric traceless tensor of rank $n$
constructed from components $x_a$, $a= 1,...,d$, of
$D$-dimensional vector $x$ and Kronecker symbols:
\begin{eqnarray}
x^{(n)}_{a_1\cdots a_n}=x_{a_1}\cdots x_{a_n}- \mbox{traces}.
\nonumber
\end{eqnarray}
By definition, the contraction  of $x^{(n)}$ with $y^{(n)}$ is
written as
\begin{eqnarray}
 x^{(n)}y^{(n)}=\sum_{k=0}^{\{\frac{n}{2}\}}d_k^{(n)}x^{2k}y^{2k}(xy)^{n-2k}
\equiv {\cal F}^{(n)}(x^2y^2,xy) \nonumber
\end{eqnarray}
where $\{n/2\}$ denotes the integer part of $n/2$, and
$d_0^{(n)}\equiv 1$. With fixing ${\cal F}^{(n)}(0,b)=b^n$ and
condition $\partial^2_x{\cal F}^{(n)}(x^2y^2,xy)=0$ the function
${\cal F}^{(n)}(a,b)$ is defined unambiguously:
\begin{eqnarray}
{\cal F}^{(n)}(a,b)=
b^n\sum_{k=0}^{\{\frac{n}{2}\}}d_k^{(n)}\left(\frac{a}{b^2}\right)^{k},
\ d_k^{(n)}= \frac{(-1)^{k}n! \Gamma(\xi+n-k-1)}{4^k
k!(n-2k)!\Gamma(\xi+n-1)}. \nonumber
\end{eqnarray}
We shell use the following notations:
\begin{eqnarray}
L(\alpha;x)\equiv \frac{1}{(x^2)^{\alpha}}, \
L^{(n)}(\alpha;x)\equiv x^{(n)}L(\alpha+n;x) \nonumber\\
\lambda_a(x;y,z)\equiv\frac{(x-y)_a}{(x-y)^2}-
\frac{(x-z)_a}{(x-z)^2}, \nonumber
\\
V(\alpha_1,\alpha_2,\alpha_3;x_1,x_2,x_3)= \nonumber
\\
=L(\Delta_{12};x_1-x_2)L(\Delta_{13};x_1-x_3)L(\Delta_{23};x_2-x_3)=
\nonumber
\\
=
(x_1-x_2)^{-2\Delta_{12}}(x_1-x_3)^{-2\Delta_{13}}(x_2-x_3)^{-2\Delta_{23}}
\nonumber
\\
V^{(n)}(\alpha,\beta,\gamma;x,y,z)= V(\alpha,\beta,\gamma;x,y,z)
\lambda^{(n)}(x;y,z).
 \nonumber
\end{eqnarray}
where $x$, $y$, $z$, $x_1$, $x_2$, $x_3$ are  vectors of
$D$-dimensional space, and
\begin{eqnarray}
\Delta_{12}= \frac{\alpha_1+\alpha_2-\alpha_3}{2}, \ \Delta_{13}=
\frac{\alpha_1+\alpha_3-\alpha_2}{2}, \ \Delta_{23}=
\frac{\alpha_2+\alpha_3-\alpha_1}{2}. \nonumber
\end{eqnarray}

 The Wilson expansion for the 4-point correlation function $W(x,y,s,t)$
of the scalar field $\Phi_\alpha$ with dimension $\alpha$ reads
 \cite{FP}, \cite{LM} :
\begin{eqnarray}
W(x,y,s,t)=\sum_{n,l}f_{ln}\int
V^{(l)}(\beta_{ln},\alpha,\alpha;z,x,y) V^{(l)}({\tilde
\beta}_{ln},\alpha,\alpha;z,s,t)dz. \label{we1}
\end{eqnarray}
Here, $f_{ln}$ are the Wilson coefficients,
$V^{(l)}(z,x,y;\beta_{ln},\alpha,\alpha)$ is (up to a constant
amplitude) the 3-point correlation function of two fields
$\Phi_\alpha$ and one symmetric traceless l-component tensor field
with dimension $d_n$, and ${\tilde \beta}_{ln}$ denotes the
"shadow" in respect to $d_{nl}$ dimension:
\begin{eqnarray}
{\tilde \beta_{ln}}\equiv 2\xi- \beta_{ln}-2l. \nonumber
\end{eqnarray}

In virtue of
$V^{(l)}(z,x,y;\beta_{ln},\alpha,\alpha)=(-1)^{l}V^{(l)}(z,x,y;\beta_{ln},\alpha,\alpha)$
and symmetry $W(x,y,s,t)=W(y,x,s,t)=W(x,s,y,t)$ we conclude that
in (\ref{we1}) the summation parameter $l$ is even, and the
crossing symmetry equation
\begin{eqnarray}
\sum_{n,l}f_{ln}\int  V^{(l)}(\beta_{ln},\alpha,\alpha;z,x,y)
V^{(l)}({\tilde \beta}_{ln},\alpha,\alpha;z,s,t) dz= \label{ks}
\\
=\sum_{n,l}f_{ln}\int  V^{(l)}(\beta_{ln},\alpha,\alpha;z,x,s)
V^{(l)}({\tilde \beta}_{ln},\alpha,\alpha;z,y,t) dz \nonumber
\end{eqnarray}
must be fulfilled. It is a non-trivial restriction on the possible
values of the Wilson coefficients and dimensions of fields. We
show how one can be expressed  in a form of exact analytical
relations not containing coordinates $x,y,s,t,z$.

\section{Crossing symmetry and sum rules}
It will be convenient for compact writing of formulas to use the
notation $\xi$ for half dimension of space $D$ and a short
notation for the product of $\Gamma$-functions:
\begin{eqnarray}
\xi\equiv \frac{D}{2}, \ \ \Gamma(a,b,\cdots,c)\equiv
\Gamma(a)\Gamma(b)\cdots\Gamma(c). \nonumber
\end{eqnarray}
If our expressions will contain the letter with prime, it will
have the following meaning:
\begin{eqnarray}
\alpha^\prime \equiv \xi-\alpha. \nonumber
\end{eqnarray}
The equality (\ref{ks}) is exact, but one is a integral equation
with infinite number of terms, and a direct analysis of them is
not easy. We obtain an evident form for the following consequence
of crossing symmetry equation
\begin{eqnarray} \sum_{n,l}f_{ln}\int
\frac{ (s-t)^{(m)}V^{(l)}(\beta_{ln},\alpha,\alpha;z,x,y)
V^{(l)}({\tilde
\beta_{ln}},\alpha,\alpha;z,s,t)}{(s-t)^{2(\gamma+m)}} dxdsdz =
\nonumber
\\
=\sum_{n,l}f_{ln}\int
\frac{(s-t)^{(m)}V^{(l)}(\beta_{ln},\alpha,\alpha;z,x,s)
V^{(l)}({\tilde
\beta}_{ln},\alpha,\alpha;z,y,t)}{(s-t)^{2(\gamma+m)}} dxds dz
\label{ks1}
\end{eqnarray}
 It is important that $(\ref{ks1})$ must be fulfilled for
 arbitrary $\gamma$ and all integer $m$.

   The first step of calculation is a direct integration over $x$.
In Appendix it is shown that with help of the formula
\begin{eqnarray}
\int L^{(n)}(\alpha;x-z)L(\beta;z-y) dz= \nonumber
\\
= \pi^\xi
\frac{\Gamma(\alpha',\beta',\xi-\alpha'-\beta'+n)}{\Gamma(\alpha+n,\beta,2\xi-\alpha-\beta)}
L^{(n)}(\alpha+\beta-\xi;x-y) \label {int}
\end{eqnarray}
one can integrate $V^{(l)}(z,x,y;\beta_{ln},\alpha,\alpha)$ over
$x$. After that the crossing symmetry equation takes the form:
\begin{eqnarray}
\sum_{n,l}f'_{ln}\int  \frac{
(s-t)^{(m)}(y-z)^{(l)}V^{(l)}({\tilde
\beta_{ln}},\alpha,\alpha;z,s,t)}{(s-t)^{2(\gamma+m)}(y-z)^{2(\alpha+l-\xi)+\beta_{ln}}}
dsdz \nonumber =
\\
=\sum_{n,l}f'_{ln}\int \frac{(s-t)^{(m)}(s-z)^{(l)}
V^{(l)}({\tilde \beta}_{ln},\alpha,\alpha; z,y,t
)}{(s-t)^{2(\gamma+m)}(s-z)^{2(\alpha+l-\xi)+\beta_{ln}}} ds dz
\label{ks2},\\
f'_{ln}= f_{ln}\pi^\xi
\frac{\Gamma(\alpha'+\beta_{ln}/2+l,-\alpha',\xi-\beta_{ln}/2)}
{\Gamma(\alpha-\beta_{ln}/2,2\xi-\alpha,\beta_{ln}/2+l)}.
\nonumber
\end{eqnarray}
 Now we do contractions of indexes in (\ref{ks2}).
For compact writing of results we use the shift operator
$T_\epsilon$ acting on functions of $\epsilon$ as follows:
\begin{eqnarray}
T_\epsilon f(\epsilon)=f(\epsilon+1).  \nonumber
\end{eqnarray}
 Let us denote $A=(\alpha_1,\alpha_2,\alpha_3)$, $X=(x_1,x_2,x_3)$,
 $E=(\epsilon_1,\epsilon_2,\epsilon_3)$, $R=(\rho_1,\rho_2,\rho_3)$,
\begin{eqnarray}
S( A;X)\equiv S(\alpha_1,\alpha_2, \alpha_3; x_1,x_2,x_3)\equiv
\label{nS}
\\
\int L(\alpha_1;x_1-y) L(\alpha_2;x_2-y)L(\alpha_3;x_3-y) dy,
\nonumber
\\
S_n( A;X)\equiv S_n(\alpha_1,\alpha_2,\alpha_3; x_1,x_2,x_3)\equiv
\label{nSn}
\\
\int L^{(n)}(\alpha_1;x_1-y)\lambda^{(n)}(y;x_2,x_3)
L(\alpha_2;x_2-y) L(\alpha_3;x_3-y) dy  \nonumber
\end{eqnarray}
It is shown in Appendix that the result of contraction of tensor
indexes  in the function $S_n( A;X)$ can be presented as
\begin{eqnarray}
S_n( A;X)= T^{(n)}(E,R) G(A,E,R)S( A+R;X)|_{E=R=0}, \label{Tn}
\end{eqnarray}
where
\begin{eqnarray}
 T^{(n)}(E,R)={\cal F}^{(n)}({\cal M}(E,R),{\cal N}(E,R)),
\nonumber\\
 {\cal M}(E,R) =T_{\epsilon_1}^{2}T_{\rho_1}[
(T_{\epsilon_2}+T_{\epsilon_3})(T_{\epsilon_2}T_{\rho_2}+T_{\epsilon_3}T_{\rho_3})
-T_{\epsilon_2}T_{\epsilon_3}T_{\rho_1}],
\nonumber\\
{\cal N}(E,R)=\frac{1}{2}T_{\epsilon_1}[
(T_{\epsilon_2}+T_{\epsilon_3})(T_{\rho_2}-T_{\rho_3})
+(T_{\epsilon_2}-T_{\epsilon_3})T_{\rho_1}]. \nonumber
\\
G(A,E,R,)= \prod_{i=1}^3\frac{\Gamma(\alpha_{nli}+\rho_i,
1-\alpha'_{nli}+\rho_i)}{\Gamma(\alpha_{nli}+\epsilon_i,
1-\alpha'_{nli})},
 \nonumber\\
\alpha_{nl1}=\alpha+ \frac{\beta_{nl}}{2}-\xi,\
\alpha_{nl2}=\alpha_{nl3}= \xi-\frac{\beta_{nl}}{2}-l. \nonumber
\end{eqnarray}
With (\ref{Tn}) the problem of integration in (\ref{ks2}) is
reduced to the case $l=0$ and is solved directly bei integration
formula (\ref{int}) (see Appendix). The final result is formulated
in the following way. The crossing symmetry relation (\ref{ks2})
is equivalent to the equality
\begin{eqnarray}
\sum_{nl}f'_{nl}\Psi_{nl}(\alpha,\beta_{nl};\gamma,m)=0
 \label{SR}
\end{eqnarray}
fulfilling for arbitrary value of parameter $\gamma$ and integer
$m\geq 0$. The function $\Psi_{nl}(\alpha,\beta_{nl};\gamma,m)$
can be written in the form
\begin{eqnarray}
\Psi_{nl}(\alpha,\beta_{nl};\gamma,m)=T^{(n)}(E,R)
Q(\alpha,\beta_{nl},E,R)\Omega_m(\alpha,\beta_{nl},\gamma,R)|_{E=R=0},
\nonumber
\end{eqnarray}
where
\begin{eqnarray}
Q(\alpha,\beta_{nl},E,R)=  \frac { \pi^{2(\xi+1)}
(-1)^{\rho_1+\rho_2} \Gamma(\alpha_{nl2}+\rho_3,
1-\alpha'_{nl2}+\rho_3) } {
\sin(\pi\alpha'_{nl1})\sin(\pi\alpha'_{nl2})\prod_{i=1}^3
 \Gamma(\alpha_{nli}+\epsilon_i, 1-\alpha'_{nli})
}, \nonumber
\end{eqnarray}

\begin{eqnarray}
\Omega_m(\alpha,\beta_{nl},\gamma,R)= \nonumber
\\=
\frac{\Gamma(\sigma'_1,\sigma'_2,\sigma'_3+m,\sigma'_4+m)}
{\Gamma(\sigma_1+m,\sigma_2+m,\sigma_3,\sigma_4)}-
\frac{\Gamma(\tau'_1,\tau'_2,\tau'_3+m,\tau'_4+m)}
{\Gamma(\tau_1+m,\tau_2+m,\tau_3,\tau_4)} \nonumber ,
\end{eqnarray}
with
\begin{eqnarray}
\sigma_{nl1}= \gamma+\alpha+\frac{\beta_{nl}}{2}-\xi+l, \
\sigma_{nl2}= \gamma+\alpha-\frac{\beta_{nl}}{2}-l+\rho_2+\rho_3,
\nonumber
\\
\sigma_{nl3}=2\xi- \alpha-\rho_2-\gamma, \
\sigma_{nl4}=3\xi+l-2\alpha-\rho_1-\rho_2 -\rho_3-\gamma
,\nonumber
\\
\tau_{nl1}=\gamma,\ \tau_{nl2}= \alpha-l+\rho_1+\rho_3+\gamma-\xi,
\nonumber
\\
\tau_{nl3}=3\xi-\frac{\beta_{nl}}{2}-\alpha-\rho_1-\gamma,
\nonumber
\\
\tau_{nl4}=2\xi+2l+\frac{\beta_{nl}}{2}-\alpha-\rho_1-\rho_2-\rho_3-\gamma.
\nonumber
\end{eqnarray}
\section{Conclusion}
   It has been shown that for scalar
Euclidean field theories at the critical point $\lambda_r =
\lambda^*$ the operators ${\cal
D}_{\alpha}^{r}(\lambda_{r},J_{r},J_\gamma)$ represent the
generators of the DTs. The functional  $W_{r}(\lambda_{r}, J_{r},
J_\gamma)$ is invariant in respect to the infinite set of the TDs
defined by ${\cal D}_{\alpha}^{r}(\lambda_{r},J_{r},J_\gamma)$ and
conformal transformations of $D$-dimensional Euclidean space. The
Ward identities (\ref{Wed}), (\ref{Confinv}) are the formal
expressions of this invariance.

The Weyl invariance is described by equation (\ref{Winv2}), where
the WT is presented with differential operator ${\cal
D}^{W}_\sigma(J,\gamma)$. For constant $\sigma$ it generates the
usual renormalization group equation, if one puts in (\ref{Wed})
$\alpha(x)=x\sigma$. With arbitrary $\sigma(x)$ (\ref{Winv2})
could be regarded as a solution of the considered in \cite{SO}
problem of local generalization of renormalization group
equations.

     The  Wilson  expansion (\ref{WE})
is included as an additional relation for constructed algebraic
structure. It was used for derivation of sum rule (\ref{SR}) which
must be fulfilled for arbitrary $\lambda$ and integer $m$. This
nontrivial condition enables to hope that (\ref{SR}) contains
essential information  about dimensions of composite operators and
Wilson coefficients.

We have obtained the following result. There is an infinite
algebraic structure corresponding to each model of conformal field
theory. It is given by the  commutation relations (\ref{crdt1}),
(\ref{crw1}), Wilson expansion formula (\ref{WE}), Ward identities
(\ref{Winv2}), (\ref{Wed}), (\ref{Confinv}) and by choosing of
logarithmic dimension ${\cal D}$ defining dimensions of sources
$\bar{d}_i^{log}$. An addition restriction follows from the sum
rules (\ref{SR}).  Thus, the problem could be to elaborate the
method of direct construction of the structure of such a kind.

\section*{Acknowlegements}
The work was supported by  Grant 03-01-00837 from Russian
Foundation for Basic Research.

\appendix
\section{Details of calculations}
We present some technical aspects of calculation methods used in
this paper. The basic formula for our integration over coordinates
of $D$-dimensional space is
\begin{eqnarray}
\int
 L(\alpha;x-z)L(\beta;z-y) dz =
 v(\alpha,\beta,\gamma)L(\gamma';x-y),
\label{A1}
 \\
\gamma=2\xi-\alpha-\beta,\ v(\alpha,\beta,\gamma)=
\pi^\xi\frac{\Gamma(\alpha^\prime,\beta^\prime,\gamma^\prime)}
{\Gamma(\alpha,\beta,\gamma)} \nonumber
\end{eqnarray}
It can be easily proven with help of Fourier transformation
\cite{V}.
 For derivatives we have:
\begin{eqnarray}
L^{(n)}(\alpha;x)=\frac{x^{(n)}}{x^{2(\alpha+n)}}=
\frac{(-1)^n\Gamma(\alpha)}{2^n\Gamma(\alpha+n)}
\partial^{(n)}_x\frac{1}{x^{2\alpha}}=
\nonumber \\
= \left(-\frac{T_\epsilon\partial_x}{2}\right)^{(n)}
\frac{\Gamma(\alpha)}{\Gamma(\alpha +\epsilon)x^{2\alpha}}
\Bigg|_{\epsilon=0}=
 \left(-\frac{T_\epsilon\partial_x}{2}\right)^{(n)}
\frac{\Gamma(\alpha)}{\Gamma(\alpha +\epsilon)}L(\alpha; x)
\Bigg|_{\epsilon=0} , \label{A2}
\end{eqnarray}
\begin{eqnarray}
(\partial^2_x)^n L(\alpha;x)=
(\partial^2_x)^n\frac{1}{x^{2\alpha}}=4^n \frac{\Gamma(\alpha +n,
n+1-\alpha^\prime)}
{\Gamma(\alpha,1-\alpha^\prime)x^{2(\alpha+n)}}= \label{A3}
\\
= (4T_\rho)^n\frac{\Gamma(\alpha +\rho, \rho+1-\alpha^\prime)}
{\Gamma(\alpha,1-\alpha^\prime)}L(\alpha+\rho;x)\Bigg|_{\rho=0}.
\nonumber
\end{eqnarray}
From (\ref{A1}), (\ref{A2}) we obtain the generalization of
(\ref{A1}):
\begin{eqnarray}
\int L^{(n)}(\alpha;x-z)L(\beta;z-y) dz =
v^{(n)}(\alpha,\beta,\gamma) L^{(n)}(\gamma';x-y)  \label{A4},
\\
v^{(n)}(\alpha,\beta,\gamma)\equiv
\pi^\xi\frac{\Gamma(\alpha',\beta',\gamma'+n)}
{\Gamma(\alpha+n,\beta,\gamma)}. \nonumber
\end{eqnarray}

For calculation of  the integral over $x$ in (\ref{ks1}) we use
the inversion operator $R$ acting on the function of the
$D$-dimensional vectors and defined as
\begin{eqnarray}
Rx\equiv \frac{x}{x^2},\ R f(x,y,\cdots,z) \equiv
f(Rx,Ry,\cdots,Rz). \nonumber
\end{eqnarray}
One has the following properties
\begin{eqnarray}
R^2=1,\ \ R \frac{1}{x^{2\alpha}}= x^{2\alpha}, \ \ R
\frac{1}{(x-y)^{2\alpha}}=
\frac{x^{2\alpha}y^{2\alpha}}{(x-y)^{2\alpha}}, \nonumber
\\
R\lambda(0;y,z)= y-z,\ \ \det\left( \frac{\partial (Rx)}{\partial
x} \right) = \frac{1}{x^{2d}}. \nonumber
\end{eqnarray}
Therefore, we obtain:
\begin{eqnarray}
\int V^{(l)}(0,x,y;\beta,\alpha,\alpha)dx = R^2\int
V^{(l)}(0,x,y;\beta,\alpha,\alpha)dx= \nonumber
\\
=R\int (x-y)^{(l)}
 V(0,Rx,Ry;\beta,\alpha,\alpha)d(Rx)= \nonumber\\
= R\int \frac{(x-y)^{(l)}}
{x^{2(d-\alpha)}y^{-2\alpha}(x-y)^{2\alpha-\beta}}dx =
\frac{v^{(l)}(\alpha-\beta/2-l, 2\xi-\alpha,\beta/2+l
)y^{(l)}}{y^{2(\alpha-\xi)+\beta}}. \nonumber
\end{eqnarray}
 Hence,
\begin{eqnarray}
\int V^{(l)}(z,x,y;\beta,\alpha,\alpha,)dx= \nonumber
\\
v^{(l)}(\alpha-\beta/2-l, 2\xi-\alpha,\beta/2+l
)L^{(l)}(\alpha-\xi+\frac{\beta}{2};y-z). \nonumber
\end{eqnarray}
Thus, after integration over $x$ in (\ref{ks1})  one obtains
(\ref{ks2}).

It follows from definition of $\lambda(x;y,z)_\mu$ and (\ref{A2})
that:
\begin{eqnarray}
\frac{\lambda(x;y,z)_\mu}{(x-y)^{2\alpha} (x-z)^{2\beta}}= \quad
\quad \quad \quad \quad \quad \nonumber
\\
\nonumber
 =D_\mu(y,z;\epsilon_1,\epsilon_2)
\frac{\Gamma(\alpha,\beta)}{\Gamma(\alpha+\epsilon_1,\beta+\epsilon_2)(x-y)^{2\alpha}
(x-z)^{2\beta}}\Bigg|_{\epsilon_1=\epsilon_2=0},
\end{eqnarray}
and
\begin{eqnarray}
\frac{\lambda^{(n)}(x;y,z)}{(x-y)^{2\alpha} (x-z)^{2\beta}}= \quad
\quad \quad \quad \quad \quad \nonumber
\\
\nonumber
 =D^{(n)}(y,z;\epsilon_1,\epsilon_2)
\frac{\Gamma(\alpha,\beta)}{\Gamma(\alpha+\epsilon_1,\beta+\epsilon_2)(x-y)^{2\alpha}
(x-z)^{2\beta}}\Bigg|_{\epsilon_1=\epsilon_2=0},
\end{eqnarray}
where
\begin{eqnarray}
D_\mu(y,z;\epsilon_1,\epsilon_2)\equiv
\frac{1}{2}\left(T_{\epsilon_1}\frac{\partial}{\partial y_\mu}
-T_{\epsilon_2}\frac{\partial}{\partial z_\mu} \right). \nonumber
\end{eqnarray}
Using (\ref{A2}) and notations  (\ref{nS}), (\ref{nSn}) we obtain
the following equality
\begin{eqnarray}
S_n( A;X)=
\left(-\frac{T_{\epsilon_1}\partial_{x_1}}{2}\right)^{(n)}D^{(n)}(x_2,x_3;\epsilon_2,\epsilon_3)
 G(A,E)S(A;X)|_{E=0}, \nonumber \\
G(A,E)=\frac{\Gamma(\alpha_1,\alpha_2,\alpha_3)}{\Gamma(\alpha_1+\epsilon_1,\alpha_2+\epsilon_2,
\alpha_3+\epsilon_3)} \nonumber
\end{eqnarray}
The function $S( A;X)$ is invariant in respect to translations,
i.e.  $S( A;x_1, x_2, x_3)=S( A;x_1+y, x_2+y, x_3+y)$. Therefore
\begin{eqnarray}
(\partial_{x_1}+\partial_{x_2}+\partial_{x_3})S( A;X)= 0,
\nonumber
\\
\partial_{x_i}\partial_{x_j}S( A;X)=
\frac{1}{2}(\partial_{x_k}^2-\partial_{x_i}^2-\partial_{x_j}^2) S(
A;X), \nonumber
\end{eqnarray}
where $i,j,k=1,2,3$ and $i\neq j, i\neq k, j\neq k$. By means of
(\ref{A3}), we obtain
\begin{eqnarray}
\left(\frac{T_{\epsilon_1}\partial_{x_1}}{2}\right)^{2k}
D^{2k}(x_2,x_3;\epsilon_2,\epsilon_3)G(A,E)S( A;X)|_{E=0}=
\nonumber
\\
=\frac{1}{4^{2k}}T_{\epsilon_1}^{2k}\partial_{x_1}^{2k}\left(T^2_{\epsilon_2}\partial_{x_2}^2
+T^2_{\epsilon_3}\partial_{x_3}^2-
2T_{\epsilon_2}T_{\epsilon_3}\partial_{x_2}\partial_{x_3}\right)^k
G(A,E)
S(A;X)|_{E=0}=\nonumber\\
= {\cal M}(E,R)^k G(A,E,R) S(A+R;X)|_{E=0, R=0}\nonumber ,
\end{eqnarray}
where  $R=(\rho_1,\rho_2,\rho_2)$,
\begin{eqnarray}
{\cal M}(E,R) =T_{\epsilon_1}^{2}T_{\rho_1}[
(T_{\epsilon_2}+T_{\epsilon_3})(T_{\epsilon_2}T_{\rho_2}+T_{\epsilon_3}T_{\rho_3})
-T_{\epsilon_2}T_{\epsilon_3}T_{\rho_1}],
\nonumber\\
G(A,E,R,)= \prod_{i=1}^3\frac{\Gamma(\alpha_i+\rho_i,
1-\alpha'_i+\rho_i)}{\Gamma(\alpha_i+\epsilon_i, 1-\alpha'_i)} .
 \nonumber
\end{eqnarray}
Analogously we obtain the relation
\begin{eqnarray}
\left(-\frac{T_{\epsilon_1}\partial_{x_1}}{2}
D(x_2,x_3;\epsilon_2,\epsilon_3)\right)^{l}G(A,E)S( A;X)|_{E=0}=
\nonumber
\\
=\frac{1}{4^{l}}T_{\epsilon_1}^{l}\left(T_{\epsilon_3}\partial_{x_1}\partial_{x_3}-T_{\epsilon_2}\partial_{x_1}\partial_{x_2}
\right)^lG(A,E)
S(A;X)|_{E=0}=\nonumber\\
= {\cal N}(E,R)^l G(A,E,R) S(A+R;X)|_{E=0, R=0}\nonumber ,
\end{eqnarray}
where
\begin{eqnarray}
{\cal N}(E,R)=\frac{1}{2}T_{\epsilon_1}[
(T_{\epsilon_2}+T_{\epsilon_3})(T_{\rho_2}-T_{\rho_3})
+(T_{\epsilon_2}-T_{\epsilon_3})T_{\rho_1}]. \nonumber
\end{eqnarray}

Thus, we have shown that the result of contraction of tensor
indexes  in the function $S_n( A;X)$ can be presented as
\begin{eqnarray}
S_n( A;X)= T^{(n)}(E,R) G(A,E,R)S( A+R;X)|_{E=R=0}, \nonumber
\end{eqnarray}
where
\begin{eqnarray}
 T^{(n)}(E,R)={\cal F}^{(n)}({\cal M}(E,R),{\cal N}(E,R)).
\nonumber
\end{eqnarray}

We can write the crossing symmetry equation (\ref{ks2}) as
\begin{eqnarray}
\sum_{n,l}f'_{nl} \int S_n(\alpha_{nl1},\alpha_{nl2},\alpha_{nl3};
y,s,t)L^{(m)}(\zeta_{nl}+\gamma;s-t)ds= \nonumber
\\
= \nonumber \sum_{n,l}f'_{nl} \int
S_n(\alpha_{nl1},\alpha_{nl2},\alpha_{nl3};
s,y,t)L^{(m)}(\gamma;s-t)L(\zeta_{nl};y-t) ds
\end{eqnarray}
with
\begin{eqnarray}
\alpha_{nl1}= \frac{\beta_{nl}}{2}-\alpha',\
\alpha_{nl2}=\alpha_{nl3}= \frac{\tilde{\beta_{nl}}}{2},\
\zeta_{nl}=\alpha-\frac{\tilde{\beta_{nl}}}{2}. \nonumber
\end{eqnarray}
By means of (\ref{A4}) we obtain
\begin{eqnarray}
\int S(\alpha_{1},\alpha_{2},\alpha_{3};
y,s,t)L^{(m)}(\zeta+\gamma;s-t)ds = \nonumber
\\
=\Phi_1(\alpha_1,\alpha_2,\alpha_3,\zeta,\gamma,m)
 L^{(m)}(\alpha_1+\alpha_2+\alpha_3+\zeta+\gamma-2\xi;t-y),
\nonumber
\end{eqnarray}
\begin{eqnarray}
\nonumber \int S(\alpha_{1},\alpha_{2},\alpha_{3};
s,y,t)L^{(m)}(\gamma;s-t)L(\zeta;s-y) ds= \nonumber\\
=\Phi_2(\alpha_1,\alpha_2,\alpha_3,\zeta,\gamma,m)
L^{(m)}(\alpha_1+\alpha_2+\alpha_3+\zeta+\gamma-2\xi;t-y),
\nonumber
\end{eqnarray}
where
\begin{eqnarray}
\Phi_1(\alpha_1,\alpha_2,\alpha_3,\zeta,\gamma,m)=
v_m(\gamma+\zeta,\alpha_2, 2\xi-\alpha_2-\gamma-\zeta)\times
\nonumber
\\
\times v_m(\gamma+\zeta+\alpha_2+\alpha_3-\xi,\alpha_1,
3\xi-\alpha_1-\alpha_2-\alpha_3-\gamma-\zeta), \nonumber
\end{eqnarray}
\begin{eqnarray}
\Phi_2(\alpha_1,\alpha_2,\alpha_3,\zeta,\gamma,m)=
v_{m}(\gamma,\alpha_1,2\xi-\alpha_1-\gamma)\times \nonumber
\\
\times
v_m(\alpha_1+\alpha_2+\gamma-\xi,\alpha_3,3\xi-\alpha_1-\alpha_2-\alpha_3-\gamma).
\nonumber
\end{eqnarray}
Let us denote
\begin{eqnarray}
\Phi(A,\zeta,\gamma,m)=\Phi(\alpha_1,\alpha_2,\alpha_3,\zeta,\gamma,m)\equiv
\Phi_1(A,\zeta,\gamma,m)-
\Phi_2(A,\zeta,\gamma,m), \nonumber\\
\Psi_{nl}(\alpha,\beta_{nl};\gamma,m)= T^{(n)}(E,R)
G(A_{nl},E,R)\Phi( A_{nl}+R,\gamma, \zeta_{nl})|_{E=R=0},
\nonumber
\end{eqnarray}
where
$A_{nl}\equiv(\alpha_{nl1},\alpha_{nl2},\alpha_{nl3})$. The
equation (\ref{ks2}) reads
\begin{eqnarray}
\sum_{nl}f'_{nl}\Psi_{nl}(\alpha,\beta_{nl};\gamma,m)=0.
 \nonumber
\end{eqnarray}
It is fulfilled for arbitrary value of parameter $\gamma$ and
integer $m\geq 0$. Using notations $\bar{E}\equiv
(\epsilon_1,\epsilon_3,\epsilon_2)$,
$\bar{R}\equiv(\rho_1,\rho_3,\rho_2)$,
\begin{eqnarray}
{\cal A}_m (\nu_1,\nu_2;\nu_3,\nu_4)\equiv
\frac{\Gamma(\nu'_1,\nu'_2,\nu'_3+m,\nu'_4+m)}
{\Gamma(\nu_1+m,\nu_2+m,\nu_3,\nu_4)},
 \nonumber
\end{eqnarray}
we write
\begin{eqnarray}
\Phi_1(A_{nl}+R,\gamma,m)
=\pi^{2\xi}\prod_{i=1}^2\frac{\Gamma(\alpha'_{nli}-\rho_i,
)}{\Gamma(\alpha_{nli}+\rho_i)} {\cal A}_m
(\sigma_{nl1},\sigma_{nl2};\sigma_{nl3},\sigma_{nl4}), \nonumber
\\
\sigma_{nl1}= \gamma+\alpha+\frac{\beta_{nl}}{2}-\xi+l, \
\sigma_{nl2}= \gamma+\alpha-\frac{\beta_{nl}}{2}-l+\rho_2+\rho_3,
\nonumber
\\
\sigma_{nl3}=2\xi- \alpha-\rho_2-\gamma, \
\sigma_{nl4}=3\xi+l-2\alpha-\rho_1-\rho_2 -\rho_3-\gamma,
\nonumber
\\
\Phi_2(A_{nl}+\bar{R},\gamma,m)
=\pi^{2\xi}\prod_{i=1}^2\frac{\Gamma(\alpha'_{nli}-\rho_i,
)}{\Gamma(\alpha_{nli}+\rho_i)} {\cal A}_m
(\tau_{nl1},\tau_{nl2};\tau_{nl3},\tau_{nl4}), \nonumber
\\
\tau_{nl1}=\gamma,\ \tau_{nl2}= \alpha-l+\rho_1+\rho_3+\gamma-\xi,
 \nonumber\\
\tau_{nl3}=3\xi-\frac{\beta_{nl}}{2}-\alpha-\rho_1-\gamma,
\nonumber
\\
\tau_{nl4}=2\xi+2l+\frac{\beta_{nl}}{2}-\alpha-\rho_1-\rho_2-\rho_3-\gamma.
\nonumber
\end{eqnarray}
Taking into account that $\Gamma(1-x,x)=\pi/\sin(\pi x)$, we
obtain
\begin{eqnarray}
G(A,E,R,)\prod_{i=1}^2\frac{\Gamma(\alpha'_{nli}-\rho_i,
)}{\Gamma(\alpha_{nli}+\rho_i)}= \nonumber
\\
=\prod_{i=1}^2\frac{\Gamma(\alpha_{nli}+\rho_i,
1-\alpha'_{nli}+\rho_i,\alpha'_{nli}-\rho_i)}{\Gamma(\alpha_{nli}+\epsilon_i,
1-\alpha'_{nli},\alpha_{nli}+\rho_i)}
\frac{\Gamma(\alpha_{nl2}+\rho_3,
1-\alpha'_{nl2}+\rho_3)}{\Gamma(\alpha_{nl2}+\epsilon_3,
1-\alpha'_{nl2})}=
 \nonumber\\
 =\pi^2\frac{\Gamma(\alpha_{nl2}+\rho_3,
1-\alpha'_{nl2}+\rho_3)}{\sin(\pi(\alpha'_{nl1}-\rho_1))\sin(\pi(\alpha'_{nl2}-\rho_2))}
\prod_{i=1}^3\frac{1}{\Gamma(\alpha_{nli}+\epsilon_i,
1-\alpha'_{nli})} .
 \nonumber
\end{eqnarray}
For integer $m$, and for even $n$
\begin{eqnarray}
\sin(\alpha+\pi m)=(-1)^m \sin(\alpha), \nonumber
\\ T^{(n)}(E,R)
G(A_{nl},E,R)\Phi_2( A_{nl}+R,\gamma, \zeta_{nl})|_{E=R=0}=
\nonumber
\\
 =T^{(n)}(E,R)
G(A_{nl},\bar{E},\bar{R})\Phi_2( A_{nl}+\bar{R},\gamma,
\zeta_{nl})|_{E=R=0}=
 \nonumber\\
 =T^{(n)}(E,R)
G(A_{nl},E,R)\Phi_2( A_{nl}+\bar{R},\gamma, \zeta_{nl})|_{E=R=0}.
 \nonumber
\end{eqnarray}
Therefore we can present the function
$\Psi_{nl}(\alpha,\beta_{nl};\gamma,m)$ as follows:
\begin{eqnarray}
\Psi_{nl}(\alpha,\beta_{nl};\gamma,m)=T^{(n)}(E,R)
Q(\alpha,\beta_{nl},E,R)\Omega_m(\alpha,\beta_{nl},\gamma,R)|_{E=R=0},
\nonumber
\end{eqnarray}
where
\begin{eqnarray}
Q(\alpha,\beta_{nl},E,R)=  \frac { \pi^{2(\xi+1)}
(-1)^{\rho_1+\rho_2} \Gamma(\alpha_{nl2}+\rho_3,
1-\alpha'_{nl2}+\rho_3) } {
\sin(\pi\alpha'_{nl1})\sin(\pi\alpha'_{nl2})\prod_{i=1}^3
 \Gamma(\alpha_{nli}+\epsilon_i, 1-\alpha'_{nli})
} \nonumber,
\\
\Omega_m(\alpha,\beta_{nl},\gamma,R)= \nonumber
\\={\cal A}_m
(\sigma_{nl1},\sigma_{nl2};\sigma_{nl3},\sigma_{nl4})- {\cal A}_m
(\tau_{nl1},\tau_{nl2};\tau_{nl3},\tau_{nl4}). \nonumber
\end{eqnarray}

\end{document}